\documentclass[12pt,aps,prd,floats,floatfix,amsmath,nofootinbib,superscriptaddress]{revtex4}
\usepackage{setspace}
\usepackage{axodraw4j}
\usepackage{pstricks}
\usepackage{graphicx}
\usepackage{bm}
\usepackage{epsfig}
\usepackage{latexsym}
\usepackage{color}
\usepackage{colordvi}
\usepackage{verbatim}
\usepackage{amssymb}
\usepackage{axodraw4j}
\usepackage{pstricks}
\usepackage{color}

\newcommand{\beq}{\begin{equation}}
\newcommand{\eeq}{\end{equation}}
\newcommand{\bea}{\begin{eqnarray}}
\newcommand{\eea}{\end{eqnarray}}

\begin{document}

\title{Superembedding Formalism and Supertwistors}

\author{Zuhair U. Khandker and Daliang Li}
\affiliation{Department of Physics, Yale University, New Haven, CT 06520}

\begin{abstract}
We establish a correspondence between superembedding and supertwistor methods for constructing $4$D $\mathcal{N}=1$ SCFT correlation functions by deriving a simple relation between tensors used in the two methods. Our discussion applies equally to $4$D CFTs by simply reducing all formulas to the $\mathcal{N}=0$ case.     
\end{abstract}

\maketitle

%%%%%%%%%%%%%%%%%%%%%%%%%%%%%%%%%%%%%%%%%%%%%%%%%%%%%%%%%%%%%%%%%%%%%%%%%%%%%%%%%%%%%%%%%%%%%%%%%%%%%%%%%%%%%%%%%%%%%%%%%%%%%%%%%%%%%%%%%%%%%%%%%%%
\section{Introduction}
\label{Introduction}

The dynamics of four-dimensional (4D) conformal field theories (CFTs) and superconformal field theories (SCFTs) are tightly constrained by the underlying conformal symmetry. However, these symmetries are obscured when correlators are expressed in terms of standard Minkowski-space coordinates. Both the embedding method and the twistor method allow for the construction of manifestly covariant correlators.\footnote{Embedding methods for CFTs go back to Refs.~\cite{Dirac:1936fq,Kastrup,Mack:1969rr,Ferrara,Siegel:1988gd}. Recent developments for constructing correlators and conformal blocks appear in Refs.~\cite{Cornalba,Weinberg:2010fx,Costa,SimmonsDuffin:2012uy}. The generalization to $4$D $\mathcal{N}=1$ SCFTs is given in Refs.~\cite{GSS,GKLS}. Twistor methods for (S)CFT correlators are detailed in Refs.~\cite{Siegel:2010yd,Siegel:2012di}.}

Both methods exploit the identification of $4$D $\mathcal{N}=1$ compactified Minkowski superspace $\overline{\mathcal{M}}$ with $\mathcal{G}$, the manifold of null two-dimensional subspaces of supertwistor space~\cite{Uhlmann,Ferber:1977qx,Siegel:1992ic,Kuzenko:2006mv,Kuzenko:2012tb} (and references therein). However, the prescription for constructing correlators in the two approaches appear to be different: $4$D superfields are mapped to embedding- or twistor-space fields with different transformation properties under the superconformal group. Consequently, the two methods utilize different tensors to build correlators. In this paper, we show that despite these differences, the two methods are equivalent due to a simple relation between the tensors appearing in each method.    

In Section~\ref{Twistor realization of Minkowski space}, we review the identification $\overline{\mathcal{M}}\cong\mathcal{G}$ and the description of $\mathcal{G}$ in both methods. In Section~\ref{Superfields}, we review superembedding/supertwistor superfields. In Section~\ref{Correlators} we show the equivalence between the two methods. We work out an example correlator in Section~\ref{Example} and conclude in Section~\ref{Conclusion}.

%%%%%%%%%%%%%%%%%%%%%%%%%%%%%%%%%%%%%%%%%%%%%%%%%%%%%%%%%%%%%%%%%%%%%%%%%%%%%%%%%%%%%%%%%%%%%%%%%%%%%%%%%%%%%%%%%%%%%%%%%%%%%%%%%%%%%%%%%%%%%%%%%%%
\section{Minkowski superspace}
\label{Twistor realization of Minkowski space}

Both the superembedding and the supertwistor method for constructing correlators exploit the following fact: $4$D compactified Minkowski superspace $\overline{\mathcal{M}}$, transforming under the $\mathcal{N}=1$ superconformal group $SU(2,2|1)$, can be identified with the manifold $\mathcal{G}$ consisting of null two-dimensional subspaces of supertwistor space $\mathbb{C}^{4|1}$~\cite{Uhlmann,Ferber:1977qx,Siegel:1992ic,Kuzenko:2006mv,Kuzenko:2012tb} (and references therein). We briefly review this identification (for a more thorough treatment, see Ref.~\cite{Kuzenko:2012tb}). 

Supertwistor space $\mathbb{C}^{4|1}$~\cite{Ferber:1977qx} consists of vectors
\begin{equation}
V_{A}=\begin{pmatrix}V_{a}\\
V^{\dot{a}}\\
V_{5}
\end{pmatrix},
\label{VA}
\end{equation}
\noindent where $V_a$ and $V^{\dot{a}}$ are bosonic while $V_5$ is fermionic. The $4$D $\mathcal{N}=1$ superconformal group $SU(2,2|1)$ ($SU$ for short) consists of $5\times 5$ supermatrices $U_A^{\phantom{A}B}$ satisfying unitarity and unimodularity.\footnote{Our notation for $SU$ follows Ref.~\cite{GSS}. Indices $A,B,\dots$ run over $A=\left\{a,\dot{a},5\right\}$, where $a,\dot{a}=1,2$ are $SL(2,\mathbb{C})$ indices, and our conventions for two-component spinors are those of Wess and Bagger~\cite{Wess:1992cp}.} $V_A$ transforms in the fundamental representation of $SU$
\begin{equation}
\delta_{SU}V_{A}=iT_{A}^{\phantom{A}B}V_{B},
\label{Fund}
\end{equation}
where $T_{A}^{\phantom{A}B}$ are the generators of $SU$. Supertwistor space is equipped with an $SU$-invariant inner product between vectors $V_A,W_A$ given by $\overline{V}^{A}W_{A} \equiv V_{\dot{A}} A^{\dot{A}A} W_{A}$, where $V_{\dot{A}} = \left( V_A \right)^\dagger$ and $A^{\dot{A}A}$ is the metric given in Eq.~(28) of Ref.~\cite{GSS}. The conjugate supertwistor $\overline{V}^A \equiv V_{\dot{A}} A^{\dot{A}A}$ is an antifundamental, $\delta_{SU}\overline{V}^{A}=-i\overline{V}^{B}T_{B}^{\phantom{B}A}$.

We define $\mathcal{G}$ to be the manifold of null two-dimensional subspaces of $\mathbb{C}^{4|1}$. We review two ways of realizing $\mathcal{G}$, which underpin the supertwistor and superembedding methods.

\subsection{`Supertwistor' description of $\mathcal{G}$}

A two-dimensional subspace of $\mathbb{C}^{4|1}$ can be spanned with two linearly-independent\footnote{In the terminology of Ref.~\cite{Buchbinder:1998qv}, the bodies of $V_A^{1,2}$ must be linearly independent.} supertwistors $V_{A}^{1,2}$. We write this pair of supertwistors as $V_{A}^{\phantom{A}\tilde{c}}$, where $\tilde{c}=1,2$. The conjugate supertwistor pair is given by $\overline{V}^{\dot{\tilde{c}}A}\equiv  \left(V_{A}^{\phantom{A}\tilde{c}}\right)^\dagger A^{\dot{A}A}$. We define $\mathcal{V}$ to be the space spanned by $\left(V_{A}^{\phantom{A}\tilde{c}}, \overline{V}^{\dot{\tilde{c}}A}\right)$.

To generate a two-dimensional null subspace, $V_{A}^{\phantom{A}\tilde{c}}$ must satisfy the null condition
\begin{equation}
\overline{V}^{\dot{\tilde{c}}A}V_{A}^{\phantom{A}\tilde{c}}=0,\hspace{10mm} \tilde{c},\dot{\tilde{c}}=1,2.
\label{Null}
\end{equation}
\noindent The choice of $V_{A}^{\phantom{A}\tilde{c}}$, however, is not unique. One could perform a rotation $V_{A}^{\prime\phantom{A}\tilde{c}}=V_{A}^{\phantom{A}\tilde{d}}g_{\tilde{d}}^{\phantom{d}\tilde{c}}$,
where $g_{\tilde{d}}^{\phantom{d}\tilde{c}}\in GL(2,\mathbb{C})$
($GL$ for short), and $V_{A}^{\prime\phantom{A}\tilde{c}}$ would
describe the same subspace. We thus view $V_{A}^{\phantom{A}\tilde{c}}$
as an object having both an $SU$ index and a $GL$ index, transforming as Eq.~(\ref{Fund}) under $SU$, but subject to the equivalence relation
\begin{equation}
V_{A}^{\phantom{A}\tilde{c}}\sim V_{A}^{\phantom{A}\tilde{d}}g_{\tilde{d}}^{\phantom{d}\tilde{c}},\hspace{10mm} g_{\tilde{d}}^{\phantom{d}\tilde{c}} \in GL.
\label{VAcEquiv}
\end{equation}
\noindent Henceforth, indices with a tilde $\left(\tilde{c},\tilde{d},\dot{\tilde{c}},\dot{\tilde{d}},... = 1,2 \right)$ will denote $GL$ indices. 
They are \emph{local} indices, transforming under changes of basis localized on a particular two-dimensional subspace, and do not transform under $SU$.

We consider the open domain of $\mathcal{G}$ where the $2\times 2$ submatrix $V_{a}^{\phantom{A}\tilde{c}}$ is non-singular, so it can be rotated into $\delta_{a}^{\phantom{a}\tilde{c}}$ by a $GL$ tranformation, and $V_{A}^{\phantom{A}\tilde{c}}$ can be written as
\begin{equation}
V_{A}^{\phantom{A}\tilde{c}}=P_{A}^{\phantom{a}\tilde{b}}\left(g_{V}\right)_{\tilde{b}}^{\phantom{b}\tilde{c}},\hspace{5mm} \text{where }  P_{A}^{\phantom{a}\tilde{b}}=\begin{pmatrix}\delta_{a}^{\phantom{a}\tilde{b}}\\
iy^{\dot{a}\tilde{b}}\\
2i\theta^{\tilde{b}}
\end{pmatrix} \text{ and } \left(g_{V}\right)_{\tilde{b}}^{\phantom{b}\tilde{c}}\in GL.
\label{P}
\end{equation}
\noindent The $\left\{P_{A}^{\phantom{a}\tilde{b}}\right\}$,
called the Poincar\'{e} section, label the equivalence classes of
$V_{A}^{\phantom{A}\tilde{c}}$. An $SU$ transformation induces a
transformation on the Poincar\'{e} section, and one finds that $\left(y,\theta\right)$ transform precisely as 
coordinates on chiral superspace. Similarly, the conjugate supertwistor pair, 
\begin{equation}
\overline{V}^{\dot{\tilde{c}}A}=\left(\bar{g}_{\overline{V}}\right)_{\phantom{c}\dot{\tilde{b}}}^{\dot{\tilde{c}}}\overline{P}^{\dot{\tilde{b}}A}, \hspace{5mm} \text{where } \overline{P}^{\dot{\tilde{b}}A}=\begin{pmatrix}-i\bar{y}^{\dot{\tilde{b}}a} & \delta_{\phantom{d}\dot{a}}^{\dot{\tilde{b}}} & -2i\bar{\theta}^{\dot{\tilde{b}}}\end{pmatrix} \text{ and } \left(\bar{g}_{\overline{V}}\right)_{\phantom{c}\dot{\tilde{b}}}^{\dot{\tilde{c}}}\in GL,
\label{Pbar}
\end{equation}
yields antichiral superspace. The null condition, Eq.~(\ref{Null}), fixes $\bar{y}^{\dot{a}\tilde{a}} = y^{\dot{a}\tilde{a}} -4i\bar{\theta}^{\dot{a}}\theta^{\tilde{a}}$, giving rise to the real coordinate $x=\frac{y+\bar{y}}{2}$ and standard $4$D Minkowski superspace $\mathcal{M}\equiv\mathbb{R}^{4|4}$. This construction is easily generalized to $\mathcal{N}$-extended Minkowski superspace (including $\mathcal{N}=0$) as reviewed in~\cite{Kuzenko:2012tb} (see also Ref.~\cite{Maio:2012tx}).

To keep track of minus signs that arise when permuting objects like $V_{A}^{\phantom{A}\tilde{c}}$, we define
\begin{eqnarray}
\sigma(AB) &\equiv& \left\{ \begin{array}{l}
-1 \text{ if } A\in\left\{ a,\dot{a}\right\} \text{ and } B\in\left\{ b,\dot{b}\right\} \\
+1 \text{ otherwise},
\end{array} \right. \\
\sigma(A) &\equiv& \sigma(AA).
\end{eqnarray}
\noindent For example, $V_{A}^{\phantom{A}\tilde{c}}V_{B}^{\phantom{A}\tilde{d}}=-\sigma(A)\sigma(B)\sigma(AB)V_{B}^{\phantom{A}\tilde{d}}V_{A}^{\phantom{A}\tilde{c}}$.

\subsection{`Superembedding' description of $\mathcal{G}$}

The $SU$-invariant inner product identifies the supertwistor $V_A$ as a one-form on the space of conjugate supertwistors. Therefore, another natural way to describe $\mathcal{G}$ is to use graded two-forms (bi-supertwistors) $X_{AB}$~\cite{Kuzenko:2012tb}. In particular, one uses $\left(V,\overline{V}\right)$ to construct
\begin{equation}
X_{AB}\equiv-\frac{i}{2}\sigma(B)V_{A}^{\phantom{a}\tilde{c}}V_{B}^{\phantom{a}\tilde{d}}\epsilon_{\tilde{c}\tilde{d}}=X^{+}\begin{pmatrix}\frac{i}{2}\epsilon_{ab} & \frac{1}{2}(y\epsilon)_{\phantom{b}a}^{\dot{b}} & \theta_{a}\\
-\frac{1}{2}(y\epsilon)_{\phantom{b}b}^{\dot{a}} & -\frac{i}{2}y^{2}\epsilon^{\dot{a}\dot{b}} & i(y\theta)^{\dot{a}}\\
\theta_{b} & i(y\theta)^{\dot{b}} & 2i\theta^{2}
\end{pmatrix} 
\label{X}
\end{equation}
\noindent and its conjugate
\begin{equation}
\overline{X}^{AB}\equiv-\frac{i}{2}\sigma(A)\overline{V}^{\dot{\tilde{c}}A}\overline{V}^{\dot{\tilde{d}}B}\epsilon_{\dot{\tilde{c}}\dot{\tilde{d}}}=\overline{X}^{+}\begin{pmatrix}-\frac{i}{2}\bar{y}^{2}\epsilon^{ab} & -\frac{1}{2}(\epsilon\bar{y})_{\phantom{a}\dot{b}}^{a} & -i(\bar{\theta}\bar{y})^{a}\\
\frac{1}{2}(\epsilon\bar{y})_{\dot{a}}^{\phantom{a}b} & \frac{i}{2}\epsilon_{\dot{a}\dot{b}} & \bar{\theta}_{\dot{a}}\\
-i(\bar{\theta}\bar{y})^{b} & \bar{\theta}_{\dot{b}} & -2i\bar{\theta}^{2}
\end{pmatrix}
\label{Xbar}
\end{equation}
\noindent where
\begin{equation}
X^{+}=g_{V}\equiv\mathrm{det}\left(g_{V}\right)_{\tilde{b}}^{\phantom{b}\tilde{c}}, \hspace{10mm} 
\overline{X}^{+}=\bar{g}_{\overline{V}}\equiv \mathrm{det}\left(\bar{g}_{\overline{V}}\right)_{\phantom{c}\dot{\tilde{d}}}^{\dot{\tilde{c}}}.
\label{X+}
\end{equation}

These definitions, together with the properties of $V$ in Eq.~(\ref{Fund}) and Eqs.~(\ref{Null}-\ref{VAcEquiv}) imply~\cite{GSS}
\begin{eqnarray}
&& X_{AB} = \sigma(AB) X_{BA} \label{XSymm} \\
&& \delta_{SU}X_{AB}=iT_{A}^{\phantom{A}A'}X_{A'B}+\sigma(AB)iT_{B}^{\phantom{B}B'}X_{B'A}.  \label{XTransf} \\
&& \overline{X}^{AB} = A^{\dot{A}A} A^{\dot{B}B} X_{\dot{A}\dot{B}}, \hspace{5mm} \text{where} \hspace{5mm} X_{\dot{A}\dot{B}} = \left(X_{BA}\right)^\dagger. 		  \label{XConj} \\
&& \left[X_{AB}X_{CD}\right]_{\mathbf{16}}=0 , \hspace{5mm} \left[\overline{X}^{AB}\overline{X}^{CD}\right]_{\overline{\mathbf{16}}} = 0, \label{16} \\
&& \left[\overline{X}^{AB}X_{BC}\right]_{\mathbf{24}} = 0, \label{24} \\
&& \left(X,\overline{X}\right) \sim \left(\lambda X,\bar{\lambda} \overline{X}\right), \hspace{5mm} \lambda \in \mathbb{C}-\left\{0\right\}. \label{XHomog} 
\end{eqnarray}
where in Eqs.~(\ref{16}) and (\ref{24}), the boldface subscripts denote the dimension of the $SU$ irreducible representation being projected onto (see Ref.~\cite{GSS} for more details). The null condition, Eq.~(\ref{Null}), implies $\bar{X}^{AB}X_{BC}=0$ (which also follows from Eqs.~(\ref{16},\ref{24})).  

%Together, the $\mathbf{16}$, $\overline{\mathbf{16}}$, and $\mathbf{24}$ constraints imply $\left[X_{AB}\overline{X}^{BA}\right]_{\mathbf{1}} = 0$, so $\bar{X}^{AB}X_{BC}=0$ for any values of $A$ and $C$. 

In Ref.~\cite{GSS}, superembedding space $\mathcal{E}$ was defined to be the space spanned by $\left(X,\overline{X}\right)$ satisfying Eqs.(\ref{XSymm}-\ref{XConj}). The constraints in Eqs.(\ref{16}-\ref{24}) and the identification in Eq.~(\ref{XHomog}) reduce $\mathcal{E}$ to $\mathcal{G}$~\cite{GSS,Kuzenko:2012tb}.

In summary, both supertwistors, $\left(V,\overline{V}\right)$, and superembedding coordinates, $\left(X,\overline{X}\right)$, can be used to describe $\mathcal{G}$. The advantage is that their $SU$ transformation rules, Eqs.~(\ref{Fund}) and (\ref{XTransf}), respectively, are linear. In the rest of this paper, we will discuss how this can be used to simplify the construction of SCFT correlators.

%%%%%%%%%%%%%%%%%%%%%%%%%%%%%%%%%%%%%%%%%%%%%%%%%%%%%%%%%%%%%%%%%%%%%%%%%%%%%%%%%%%%%%%%%%%%%%%%%%%%%%%%%%%%%%%%%%%%%%%%%%%%%%%%%%%%%%%%%%%%%%%%%%%
\section{Superfields}
\label{Superfields}

Manifestly covariant correlators can be constructed for fields on $\mathcal{E}$ or $\mathcal{V}$. To map results back to $\mathcal{M}$ requires a correspondence between these fields and standard superfields on $\mathcal{M}$~\cite{GKLS,Siegel:2012di}, which we now review.

A generic primary superfield $\phi$ on $\mathcal{M}$ is specified by the quantum numbers $\left(j,\bar{j},q,\bar{q}\right)$~\cite{Osborn:1998qu}, where $\left(j,\bar{j}\right)$
are its $SL(2,\mathbb{C})$ Lorentz quantum numbers and $q,\bar{q}$ are related to the
scaling dimension $\Delta$ and $U(1)_R$ charge of its lowest component field by
\begin{equation}
q\equiv\frac{1}{2}\left(\Delta+\frac{3}{2}R\right), \hspace{5mm} \bar{q}\equiv\frac{1}{2}\left(\Delta-\frac{3}{2}R\right).
\end{equation}
\noindent We illustrate the procedure for uplifting $\phi_{c}\sim\left(\frac{1}{2},0,q,\bar{q}\right)$ to a superfield $\Phi$ on either $\mathcal{E}$ or $\mathcal{V}$. 

\bigskip
\bigskip
\noindent $\underline{\text{Superembedding Method}}$
\begin{equation}
\Phi_{A}\left(X,\overline{X}\right)=\left(X^{+}\right)^{-\left(q+\frac{1}{2}\right)}\left(\overline{X}^{+}\right)^{-\bar{q}}\cdot2iX_{A}^{\phantom{A}c}\phi_{c}\left(x,\theta,\bar{\theta}\right)
\label{IaDefn}
\end{equation}
\begin{equation}
\Phi_{A}\left(\lambda X,\bar{\lambda}\overline{X}\right)=\lambda^{-\left(q-\frac{1}{2}\right)}\bar{\lambda}^{-\bar{q}}\Phi_{A}\left(X,\overline{X}\right)
\label{IaHomog}
\end{equation}
Here, $\Phi_{A}$ is an $SU$ fundamental, this time with first four components fermionic and fifth component bosonic. 
Note that essentially $X_{A}^{\phantom{A}c}$ is being used as a vierbein in Eq.~(\ref{IaDefn}) to convert the Lorentz spinor index on $\phi_c$ to an $SU$ index on $\Phi_A$. To
recover $\phi_{c}$, we plug the relation $2iX_{a}^{\phantom{A}c}=X^{+}\delta_{a}^{\phantom{a}c}$ into Eq.~(\ref{IaDefn}) and obtain, 
\begin{equation}
\phi_{a} = \left. \Phi_{A=a} \right|_{X^+,\overline{X}^+=1}.
\label{IaProj}
\end{equation}
Given a correlator of $\Phi_A$, one simply uses Eq.~(\ref{IaProj}) to recover the correlator of $\phi_a$. 

\bigskip
\bigskip
\noindent $\underline{\text{Supertwistor Method}}$
\begin{equation}
\Phi^{\tilde{c}}\left(V,\overline{V}\right)=\left(g_{V}\right)^{-\left(q+\frac{1}{2}\right)}\left(\bar{g}_{\overline{V}}\right)^{-\bar{q}}\phi^{a}\left(x,\theta,\bar{\theta}\right)V_{a}^{\phantom{a}\tilde{c}}
\label{IIDefn}
\end{equation}
\begin{equation}
\Phi^{\tilde{c}}\left(\lambda V,\bar{\lambda}\overline{V}\right)=\lambda^{-2q}\bar{\lambda}^{-2\bar{q}}\Phi^{\tilde{c}}\left(V,\overline{V}\right)
\label{IIHomog}
\end{equation}
\noindent Here, $\Phi^{\tilde{c}}$ is an $SU$ singlet; it only has a $GL$ index. On the right-hand-side of Eq.~(\ref{IIDefn}) we see that $V_{a}^{\phantom{a}\tilde{c}}$ is being used as a vierbein to convert the Lorentz spinor index on $\phi^a$ to the $GL$ index on $\Phi^{\tilde{c}}$~\cite{Siegel:2012di}. Given a correlator of $\Phi^{\tilde{c}}$, one `peels off' factors of $V_{a}^{\phantom{a}\tilde{c}}$ to recover the correlator of $\phi^a$~\cite{Siegel:2012di}.    

\bigskip
\bigskip

The verification of the above correspondence and the generalization to arbitrary $\left(j,\bar{j},q,\bar{q}\right)$ is given in Refs.~\cite{GKLS,Siegel:2012di}.

%%%%%%%%%%%%%%%%%%%%%%%%%%%%%%%%%%%%%%%%%%%%%%%%%%%%%%%%%%%%%%%%%%%%%%%%%%%%%%%%%%%%%%%%%%%%%%%%%%%%%%%%%%%%%%%%%%%%%%%%%%%%%%%%%%%%%%%%%%%%%%%%%%%
\section{Tensors and Invariants}
\label{Correlators}

In the superembedding method, correlation functions of superfields on $\mathcal{E}$ are built from $X$ and $\overline{X}$. Any tensor constructed from these coordinates has at least two indices, and we utilize the notation of Ref.~\cite{GKLS} for two-index tensors,
\begin{equation}
\left(1 \bar{2} 3 \cdots \bar{N}\right)_A^{\phantom{A}B} \equiv \left(X_1 \sigma \bar{X}_2 X_3\sigma \cdots \bar{X}_N \right)_A^{\phantom{A}B},
\label{XTensor}
\end{equation}
with the obvious analogs for $\left(1 \bar{2} \cdots N\right)_{AB}$ and $\left(\bar{1} 2 \cdots \bar{N}\right)^{AB}$. 

In the supertwistor method, superfields on $\mathcal{V}$ only have $GL$ indices, so the building block for correlators is obtained by contracting the $SU$ indices of a $\overline{V}_i^{\dot{\tilde{c}}A}$ and a $V_{jA}^{\phantom{ja}\tilde{c}}$ to form~\cite{Siegel:2012di}
\begin{equation}
Y_{\bar{i}j}^{\dot{\tilde{c}}\tilde{c}} \equiv \frac{1}{2} \overline{V}_{i}^{\dot{\tilde{c}}A}V_{jA}^{\phantom{ja}\tilde{c}}.
\label{Y}
\end{equation}
Note that $Y_{\bar{i}i}^{\dot{\tilde{c}}\tilde{c}}=0$
because of Eq.~(\ref{Null}). Also note that $Y_{\bar{i}j}^{\dot{\tilde{c}}\tilde{c}}$ does not transform under $SU$ and is a superconformal invariant.
Since a $GL$ index is a local index attached to a particular $V_i$, it can only be contracted with another $GL$ index attached to the same $V_i$. Thus, two-index tensors built from $Y_{\bar{i}j}^{\dot{\tilde{c}}\tilde{c}}$ can be written as 
\begin{equation}
Y_{\bar{1}2\bar{3}\cdots N}^{\dot{\tilde{c}}\tilde{c}} \equiv Y_{\bar{1}2}^{\dot{\tilde{c}}\tilde{a}_1} \epsilon_{\tilde{a}_1\tilde{a}_{2}}  Y_{\bar{3}2}^{\dot{\tilde{a}}_2\tilde{a}_2} \epsilon_{\dot{\tilde{a}}_2\dot{\tilde{a}}_3} Y_{\bar{3}4}^{\dot{\tilde{a}}_3\tilde{a}_3}\cdots Y_{\overline{N-1}N}^{\dot{\tilde{a}}_{N-1}\tilde{c}} 
\label{YTensor}
\end{equation}
with the obvious analogs for $Y_{\bar{1}2\bar{3}\cdots \bar{N}}^{\dot{\tilde{a}}\dot{\tilde{b}}}$ and $Y_{1\bar{2}3\cdots N}^{\tilde{a}\tilde{b}}$.

The equivalence of the superembedding and supertwistor methods stems from the following simple correspondence between the tensors in Eqs.~(\ref{XTensor}) and (\ref{YTensor}): 
\begin{eqnarray}
Y_{\bar{1}\cdots N}^{\dot{\tilde{a}}\tilde{a}} &=& \frac{2}{\bar{g}_1 g_N} \bar{g}^{\dot{\tilde{a}}}_{1\phantom{a}\dot{c}} \left(\bar{1}\cdots N \right)^{\dot{c}c} g_{N c}^{\phantom{NC}\tilde{a}}, \label{TensorReln} \\
\left(1\cdots \bar{N}\right)_A^{\phantom{A}B} &=& \frac{1}{2} V_{1A}^{\phantom{1A}\tilde{c}} \left( Y_{\bar{N}\cdots 1}^{\dot{\tilde{d}}\tilde{d}}\right) \overline{V}_N^{\phantom{N}\dot{\tilde{c}}B} \epsilon_{\tilde{c}\tilde{d}} \epsilon_{\dot{\tilde{d}}\dot{\tilde{c}}} \label{TensorReln2}
\end{eqnarray}
%
%with similar relations for $Y_{\bar{1}2\bar{3}\cdots \bar{N}}^{\dot{\tilde{a}}\dot{\tilde{b}}}$ and $Y_{1\bar{2}3\cdots N}^{\tilde{a}\tilde{b}}$. 
This is simply because when $GL$ indices on two $V$'s are contracted as in Eq.~(\ref{YTensor}), they form an $X$ according to Eq.~(\ref{X}), similiarly for $\bar{V}$ and $\bar{X}$. To derive Eq.~(\ref{TensorReln}) explicitly, one first observes that
\begin{equation}
P_{A}^{\phantom{A}(\tilde{c}=c)}=\frac{2i}{X^{+}}X_{A}^{\phantom{A}c}, \hspace{10mm} \overline{P}^{(\dot{\tilde{c}}=\dot{c})A}=-\frac{2i}{\overline{X}^{+}}\overline{X}^{\dot{c}A},
\label{PXReln}
\end{equation}
\noindent which imply
\begin{equation}
X_{AB}=\frac{2i}{X^{+}}\sigma(B)X_{A}^{\phantom{A}c}X_{B}^{\phantom{B}d}\epsilon_{cd}, \hspace{10mm}
\overline{X}^{AB}=\frac{2i}{\overline{X}^{+}}\sigma(A)\overline{X}^{\dot{c}A}\overline{X}^{\dot{d}B}\epsilon_{\dot{c}\dot{d}}.
\label{XXReln}
\end{equation}
Eqs.~(\ref{PXReln}) and (\ref{XXReln}) imply Eq.~(\ref{TensorReln}). Meanwhile, Eq.~(\ref{TensorReln2}) follows from Eqs.~(\ref{X},\ref{Xbar}). It follows that scalar invariants in both approaches are identical,
\begin{equation}
\left\langle \bar{1}\cdots N\right\rangle \equiv - \bar{1}^A \cdots N_A  = - Y_{\bar{1}\cdots N}^{\dot{\tilde{a}}\tilde{a}} Y_{\bar{1}N}^{\dot{\tilde{b}}\tilde{b}} \epsilon_{\dot{\tilde{a}}\dot{\tilde{b}}} \epsilon_{\tilde{a}\tilde{b}}.
\end{equation}

Both the superembedding and supertwistor methods reduce the construction of correlators to the task of enumerating tensors. Because of the correspondence in Eqs.~(\ref{TensorReln},\ref{TensorReln2}) this task becomes equivalent in the two methods. 
The relation between the two methods can also be seen from Eqs.~(\ref{IaDefn}) and (\ref{IIDefn}), which along with Eq.~(\ref{X}) imply that projections of superfields defined on $\mathcal{E}$ are related to superfields defined on $\mathcal{V}$ by $\Phi_{A=a} = \left(g_V\right)_a^{\phantom{a}\tilde{c}} \Phi_{\tilde{c}}$.

%%%%%%%%%%%%%%%%%%%%%%%%%%%%%%%%%%%%%%%%%%%%%%%%%%%%%%%%%%%%%%%%%%%%%%%%%%%%%%%%%%%%%%%%%%%%%%%%%%%%%%%%%%%%%%%%%%%%%%%%%%%%%%%%%%%%%%%%%%%%%%%%%%%
\section{Example}
\label{Example}
Four-dimensional $\mathcal{N}=1$ SCFT two- and three-point functions were worked out previously~\cite{Molotkov:1975gf,Osborn:1998qu,Park:1997bq}. 
%These works did not use superembedding or supertwistor methods, but the underlying philosophy was the same: find combinations of Minkowski superspace coordinates that have simpler transformation properties under $SU$. The superembedding and supertwistor methods allow one to work with manifest $SU$ covariance throughout.
Manifestly covariant two-point functions involving superfields in arbitrary $SU$ representations were worked out using the superembedding and supertwistor approaches in Ref.~\cite{GKLS} and Ref.~\cite{Siegel:2012di}, respectively.\footnote{Ref.~\cite{GKLS} made use of a variant of the index-free formalism introduced in Ref.~\cite{Costa}.} Additionally, Ref.~\cite{GKLS} worked out manifestly covariant expressions for three-point functions involving conserved current superfields. 

%Rather than repeat these examples, 
We consider here another correlator, the three-point function involving a superfield $\mathcal{T}_{c\dot{c}}(x_{1},\theta_{1},\bar{\theta}_{1})\sim(\frac{1}{2},\frac{1}{2},\frac{3}{2},\frac{3}{2})$
\footnote{$\mathcal{T}_{c\dot{c}}$ could be the supercurrent superfield~\cite{Ferrara:1974pz},
which contains the energy-momentum tensor, if it additionally satisfies $D^{c}\mathcal{T}_{c\dot{c}}=\bar{D}^{\dot{c}}\mathcal{T}_{c\dot{c}}=0$,
where $D^{c}$ is the $\mathcal{N}=1$ Poincar\'{e} super-covariant derivative.}, a chiral scalar superfield $\phi(y_{2},\theta_{2})\sim(0,0,q_{2,}0)$, and an antichiral scalar superfield $\bar{\phi}(\bar{y}_{3},\bar{\theta}_{3})\sim(0,0,0,\bar{q}_{3})$, 
\begin{equation}
\left\langle \mathcal{T}_{c\dot{c}}(x_{1},\theta_{1},\bar{\theta}_{1})\phi(y_{2},\theta_{2})\bar{\phi}(\bar{y}_{3},\bar{\theta}_{3})\right\rangle.
\end{equation}
%
%This correlator is simple and illustrates the operational equivalence between the superembedding and supertwistor approaches. 

We start with the superembedding approach. First, we introduce superfields on $\mathcal{E}$ related to the superfields on $\mathcal{M}$ by
\begin{equation}
\mathcal{T}_{A}^{\phantom{A}B}(X_{1},\bar{X}_{1})=\left(X_{1}^{+}\right)^{-2}\left(\bar{X}_{1}^{+}\right)^{-2}X_{1A}^{\phantom{1A}c}\mathcal{T}_{c\dot{c}}\bar{X}_{1}^{\dot{c}B}, \hspace{4mm}
\Phi(X_{2})=\left(X_{2}^{+}\right)^{-q_{2}}\phi, \hspace{4mm}
\bar{\Phi}(\bar{X}_{3})=\left(\bar{X}_{3}^{+}\right)^{-\bar{q}_{3}}\bar{\phi},
\end{equation}
satisfying
\begin{equation}
\mathcal{T}_{A}^{\phantom{A}B}(\lambda X_{1},\bar{\lambda}\bar{X}_{1})=\lambda^{-1}\bar{\lambda}^{-1}\mathcal{T}_{A}^{\phantom{A}B}(X_{1},\bar{X}_{1}), \hspace{5mm}
\Phi(\lambda X_{2})=\lambda^{-q_{2}}\Phi(X_{2}), \hspace{5mm}
\bar{\Phi}(\bar{\lambda}\bar{X}_{3})=\bar{\lambda}^{-\bar{q}_{3}}\bar{\Phi}(\bar{X}_{3}),
\end{equation}
\noindent and consider the correlator 
\begin{equation}
\left\langle \mathcal{T}_{A}^{\phantom{A}B}(X_{1},\bar{X}_{1})\Phi(X_{2})\bar{\Phi}(\bar{X}_{3})\right\rangle = \left(X_{1}^{+}\right)^{-2}\left(\bar{X}_{1}^{+}\right)^{-2}(X_{2}^{+})^{-q_{2}}(\bar{X}_{3}^{+})^{-\bar{q}_{3}} \left\langle X_{1A}^{\phantom{1A}c'}\mathcal{T}_{c'\dot{c}'}\bar{X}_{1}^{\dot{c}'B}\phi\bar{\phi} \right\rangle.
\end{equation}
\noindent The RHS dictates that the index $A$ attach to $X_{1}$
and the index $B$ attach to $\bar{X}_{1}$. The only possible such
tensor is $\left(1\bar{3}2\bar{1}\right)_A^{\phantom{A}B}$. Then by homogeneity, up to an overall constant $C$,
\begin{equation}
\langle\mathcal{T}_{A}^{\phantom{A}B}(X_{1},\bar{X}_{1})\Phi(X_{2})\bar{\Phi}(\bar{X}_{3})\rangle = C \frac{\left(1\bar{3}2\bar{1}\right)_A^{\phantom{A}B}}{\langle1\bar{3}\rangle^{2}\langle2\bar{1}\rangle^{2}\langle2\bar{3}\rangle^{q_{2}-1}}\delta_{q_{2},\bar{q}_{3}}
\end{equation}
\noindent To recover the $4$D correlator, one simply uses Eq.~(\ref{IaProj}),
\begin{eqnarray}
&& \left\langle  \mathcal{T}_{c\dot{c}}(x_{1},\theta_{1},\bar{\theta}_{1})\phi(y_{2},\theta_{2})\bar{\phi}(\bar{y}_{3},\bar{\theta}_{3}) \right\rangle = C \left.\frac{\left(1\bar{3}2\bar{1}\right)_{A=c}^{\phantom{A=c}B=\dot{c}}}{\langle1\bar{3}\rangle^{2}\langle2\bar{1}\rangle^{2}\langle2\bar{3}\rangle^{q_{2}-1}}\right|_{(X_{i}^{+}=1)}\delta_{q_{2},\bar{q}_{3}} \nonumber \\
&& \hspace{10mm} =
C^\prime \frac{\left(y_{1\bar{3}}+4i\theta_{1}\bar{\theta}_{3}\right)_{c\dot{d}}\left(y_{\bar{3}2}+4i\bar{\theta}_{3}\theta_{2}\right)^{\dot{d}d}\left(y_{2\bar{1}}+4i\theta_{2}\bar{\theta}_{1}\right)_{d\dot{c}}}{\left[\left(y_{\bar{3}1}+2i\theta_1\sigma\bar{\theta}_3\right)^2\right]^2\left[\left(y_{\bar{1}2}+2i\theta_2\sigma\bar{\theta}_1\right)^2\right]^2\left[\left(y_{\bar{3}2}+2i\theta_2\sigma\bar{\theta}_3\right)^2\right]^{q_2-1}} \, \delta_{q_{2},\bar{q}_{3}},
\label{Example1Ans}
\end{eqnarray}
in agreement with Ref.~\cite{Osborn:1998qu}.

We now consider the supertwistor method. The superfields on $\mathcal{V}$ are given by
\begin{equation}
T^{\dot{\tilde{c}}\tilde{c}}(V_{1},\bar{V}_{1})=g_{1}^{-2}\bar{g}_{1}^{-2}\bar{g}_{1\phantom{c}\dot{a}}^{\dot{\tilde{c}}}\mathcal{T}^{\dot{a}a}g_{1a}^{\phantom{1a}\tilde{c}}, \hspace{5mm}
\Phi(V_{2})=g_{2}^{-q_{2}}\phi, \hspace{5mm}
\bar{\Phi}(\bar{V}_{3})=\bar{g}_{3}^{-\bar{q}_{3}}\bar{\phi}
\end{equation}
\begin{equation}
T^{\dot{\tilde{c}}\tilde{c}}(\lambda V_{1},\bar{\lambda}\bar{V}_{1})=\lambda^{-3}\bar{\lambda}^{-3}T^{\dot{\tilde{c}}\tilde{c}}(V_{1},\bar{V}_{1}), \hspace{5mm}
\Phi(\lambda V_{2})=\lambda^{-2q_{2}}\Phi(V_{2}), \hspace{5mm}
\bar{\Phi}(\bar{\lambda}\bar{V}_{3})=\bar{\lambda}^{-2\bar{q}_{3}}\bar{\Phi}(\bar{V}_{3})
\end{equation}
\noindent and we consider the correlator
\begin{equation}
\left\langle T^{\dot{\tilde{c}}\tilde{c}}(V_{1},\bar{V}_{1})\Phi(V_{2})\bar{\Phi}(\bar{V}_{3})\right\rangle = g_{1}^{-2}\bar{g}_{1}^{-2}g_{2}^{-q_{2}}\bar{g}_{3}^{-\bar{q}_{3}}\bar{g}_{1\phantom{c}\dot{a}}^{\dot{\tilde{c}}} \left\langle \mathcal{T}^{\dot{a}a}\phi\bar{\phi}\right\rangle g_{1a}^{\phantom{1a}\tilde{c}}
\label{IICorr}
\end{equation}
\noindent The building blocks for this correlator are $Y_{\bar{1}2}^{\dot{\tilde{a}}\tilde{a}}$,
$Y_{\bar{3}1}^{\dot{\tilde{a}}\tilde{a}}$, and $Y_{\bar{3}2}^{\dot{\tilde{a}}\tilde{a}}$.
Just as in the superembedding approach, the RHS dictates that the index $\dot{\tilde{c}}$
attach to $\overline{V}_{1}$ while $\tilde{c}$ attach to $V_{1}$.
Thus the tensor representing this correlator must start with $Y_{\bar{1}2}^{\dot{\tilde{c}}\tilde{b}}$
and end with $Y_{\bar{3}1}^{\dot{\tilde{b}}\tilde{c}}$. The only possibility is $Y_{\bar{1}2\bar{3}1}^{\dot{\tilde{c}}\tilde{c}}$, which is related to $\left(1\bar{3}2\bar{1}\right)_A^{\phantom{A}B}$ by Eqs.~(\ref{TensorReln},\ref{TensorReln2}).
Then by homogeneity
\begin{equation}
\left\langle T^{\dot{\tilde{c}}\tilde{c}}(V_{1},\bar{V}_{1})\Phi(V_{2})\bar{\Phi}(\bar{V}_{3})\right\rangle = C^{\prime\prime} \frac{1}{\langle1\bar{3}\rangle^{2}\langle2\bar{1}\rangle^{2}\langle2\bar{3}\rangle^{q_{2}-1}}\delta_{q_{2},\bar{q}_{3}} Y_{\bar{1}2\bar{3}1}^{\dot{\tilde{c}}\tilde{c}}.
\end{equation}
\noindent Using Eq.~(\ref{TensorReln}) and comparing with Eq.~(\ref{IICorr}) gives exactly Eq.~(\ref{Example1Ans}). 

More complicated correlators involve the enumeration of more than one tensor. Eqs.~(\ref{TensorReln},\ref{TensorReln2}) imply that this procedure is equivalent in both methods. 
%An advantage of the superembedding method is that contraction of $GL$ indices is already done, and the $GL$ matrices $g_{a}^{\phantom{a}\tilde{c}}$ do not appear in intermediate steps, while an advantage of the supertwistor method is that tensors have two-component $GL$ indices rather than five-component $SU$ indices. 

%Thus the only slight difference between the superembedding and supertwistor approaches is that in the former, any contraction of $GL$ indices is already done, and the $GL$ matrices $g_{a}^{\phantom{a}\tilde{c}}$ do not appear in intermediate steps.

%%%%%%%%%%%%%%%%%%%%%%%%%%%%%%%%%%%%%%%%%%%%%%%%%%%%%%%%%%%%%%%%%%%%%%%%%%%%%%%%%%%%%%%%%%%%%%%%%%%%%%%%%%%%%%%%%%%%%%%%%%%%%%%%%%%%%%%%%%%%%%%%%%%
\section{Conclusion}
\label{Conclusion}

%In this paper, we have reviewed the superembedding and supertwistor approaches for constructing $4D$ $\mathcal{N}=1$ SCFT (and $4D$ CFT) correlators. 
We have shown an equivalence between the superembedding and supertwistor methods for constructing $4$D $\mathcal{N}=1$ SCFT (and $4$D CFT) correlators. For other applications, one of the methods may be more natural or efficient, so it seems worthwhile to have both in one's toolbag. Overall, we hope that the simplifications afforded by these approaches will eventually lead to new results in SCFTs. 

Thus far, correlators constructed using the superembedding and supertwistor approaches have been relatively simple. To construct more complicated correlators, it is imperative to have a better understanding of tensors. When constructing a correlation function, any tensor with the right index structure and homogeneity properties can appear but not all of them are linearly independent. We need a systematic way to identify such a linearly independent subset. We hope to explore this question further.

\section*{Acknowledgements}
We thank Walter Goldberger and Witold Skiba for comments on the manuscript.  This work is supported in part by DOE grant DE-FG-02-92ER40704.

%%%%%%%%%%%%%%%%%%%%%%%%%%%%%%%%%%%%%%%%%%%%%%%%%%%%%%%%%%%%%%%%%%%%%%%%%%%%%%%%%%%%%%%%%%%%%%%%%%%%%%%%%%%%%%%%%%%%%%%%%%%%%%%%%%%%%%%%%%%%%

\end{document}